# Streaming and Sublinear Approximation of Entropy and Information Distances

Sudipto Guha     Andrew McGregor*     Suresh Venkatasubramanian[†]

September 25, 2018

**Abstract**

In many problems in data mining and machine learning, data items that need to be clustered or classified are not points in a high-dimensional space, but are distributions (points on a high dimensional simplex). For distributions, natural measures of distance are not the $\ell_p$ norms and variants, but information-theoretic measures like the Kullback-Leibler distance, the Hellinger distance, and others. Efficient estimation of these distances is a key component in algorithms for manipulating distributions. Thus, sublinear resource constraints, either in time (property testing) or space (streaming) are crucial.

In this paper we design streaming and sublinear time property testing algorithms for entropy and various information theoretic distances. We start by resolving two open questions regarding property testing of distributions. Firstly, we show a tight bound for estimating bounded, symmetric *f-divergences* between distributions in a general property testing (sublinear time) framework (the so-called *combined oracle model*). This yields optimal algorithms for estimating such well known distances as the Jensen-Shannon divergence and the Hellinger distance. Secondly, we close a $(\log n)/H$ gap between upper and lower bounds for estimating entropy $H$ in this model. We provide an optimal algorithm over all values of the entropy, and for small entropy the improvement is significant. In a stream setting (sublinear space), we give the first algorithm for estimating the entropy of a distribution. Our algorithm runs in polylogarithmic space and yields an asymptotic constant factor approximation scheme. An integral part of the algorithm is an interesting use of $F_0$ (the number of distinct elements) estimation algorithms; we also provide other results along the space/time/approximation tradeoff curve.

Our results have interesting structural implications that connect sublinear time- and space-constrained algorithms. The mediating model is the random order streaming model, which assumes the input is a random permutation of a multiset and was first considered by Munro and Patterson. We show that any property testing algorithm in the combined oracle model for permutation invariant functions can be simulated in the random order model in a single pass. This addresses a question raised by Feigenbaum et al regarding the relationship between property testing and stream algorithms. Further we give a polylog-space PTAS for the estimating the entropy of a one pass random order stream. This bound cannot be achieved in the combined oracle model.

---

*Department of Computer Information Sciences, University of Pennsylvania, 3330 Walnut St, Philadelphia, 19104. **Email:** {sudipto,andrewm}@cis.upenn.edu

[†]AT&T Research Labs, 180 Park Ave, Florham Park, NJ 07928, **Email:** suresh@research.att.com

# 1 Introduction

There are many settings where the natural unit of data, rather than being a point in a high dimensional vector space, is a distribution defined on $n$ items. Examples include soft clustering [36], where the membership of a point in a cluster is described by a distribution, and anomaly detection [30], where the distance between two empirical distributions is used to detect anomalies. Typically, such settings involve large data sets, and so a natural requirement is that algorithms use small amounts of resources (space or time).

In this paper, we examine sublinear algorithms for estimating properties of distributions. On the one hand we study the complexity of estimating information theoretic distances and measures on distributions, e.g., entropy, Jensen-Shannon divergence, Hellinger and Triangular distances, to name a few, and on the other, we explore the connections between various models in sublinear algorithms, e.g., property testing models, and data streams. We discuss both these aspects below. We will not be able to review the extensive literature on either of these topics; however several good surveys, e.g., by Ron [35], Babcock et al [4] and Muthukrishnan [34], exist.

## 1.1 Problems

When dealing with distributions, distances arising from information-theoretic considerations are often more natural than distances based on $\ell_p$ norms and the like. In the first half of the paper we focus on the Ali-Silvey distances or $f$-divergences, discovered independently by Csiszár [18], and Ali and Silvey [1]. The class of $f$-divergences include many commonly used information theoretic distances, e.g., the (asymmetric) Kullback-Liebler (KL) divergence[1] and its symmetrization the Jensen-Shannon (JS) divergence, Matsusita's Divergence or the squared Hellinger distance, the (asymmetric) $\chi^2$ distance and its symmetrization, the Triangle distance. In fact for any convex function $f$ we can define the $f$-divergence $D_f(q,p) = \sum_{x \in \Omega} p(x) f(q(x)/p(x))$ if $f(1) = 0$ and $f$ is strictly convex at 1.[2]

Results of Csiszár [19], Liese and Vajda [31], Amari [3] and many others show that $f$-divergences are the unique class of distances on distributions that arise from a fairly simple set of axioms, e.g., permutation invariance, non-decreasing projections, certain direct sum theorems etc., in much the same way that $\ell_2$ is a natural measure for points in $R^n$. Moreover, all of these distances are related to each other (via the Fisher information matrix) [13] in a way that other plausible measures (most notably $\ell_2$) are not. In addition, the log-likelihood ratio $\ln \frac{q(x)}{p(x)}$ is a crucial parameter in Neyman-Pearson style hypothesis testing [17], and distances based on this (like the KL-distance and the JS-distance) appear as exponents of error probabilities for optimal classifiers. Recently, these distance measures have been used in more algorithmic contexts, as natural distances for clustering distributional data [36, 21, 5]. Batu et al [12] gave algorithms for testing closeness of distributions for the $\ell_1$ and $\ell_2$ distances, and raised the question of testing closeness of distributions under the JS-divergence.

In this paper we provide optimal (upto constants) algorithms for testing $f$-divergences of distributions. We consider the problem of estimating the entropy $H$ of a distribution, providing optimal (upto constants) upper bounds for testing entropy. This improves the previous result of Batu et al [11] by a factor $\frac{\log n}{H(p)}$. Entropy is naturally related to the JS-divergence since $JS(p,q) = \ln 2(2H((p+q)/2) - H(p) - H(q))$ where $(p+q)/2$ is the average of the two distributions.

Switching from sublinear time to sublinear space, we then focus on the streaming model and develop a one pass polylogarithmic space PTAS for estimating entropy in the random streams model, which assumes that the input is a random permutation of some fixed multiset. We subsequently derive several (regular) streaming algorithms that give a three way tradeoff between space, approximation, and number of passes. We note that these algorithms naturally imply (weaker) tradeoffs in JS distance and omit further discussion.

---
[1] Many of the measures we consider in this paper are not metrics – and several authors use constant multiples of the definitions in this paper. Traditionally, the term 'divergence' has been used to distinguish such measures from distances and metrics. We will use the terms 'distance' and 'divergence' interchangeably; a distance is not a metric unless explicitly mentioned.

[2] We can easily verify that $f(u) = u \ln u$ gives us the KL divergence; $f(u) = (\ln(2/(1+u)) + u \ln(2u/(1+u)))$ gives us the Jensen-Shannon (JS) divergence. The asymmetric Pearson's $\chi^2$ distance is realized with $f(u) = (u-1)^2$ and is symmetrized to the Triangle distance with $f(u) = (u-1)^2/(u+1)$. Matsusita's Divergence or the (squared) Hellinger distance has $f(u) = (\sqrt{u}-1)^2$. The $\ell_1$ or variational distance is realized with $f(u) = |u-1|$.

## 1.2 Models

As it turns out, sublinear algorithms for testing distributions reveal interesting structure about the relationship between property testing and stream algorithms. Feigenbaum et al [22] considered the problem of property testing in a data stream model. They showed that there exist functions (e.g., SORTED-SUPERSET, a variant of permutation, [22]) that are easy in the property testing model but hard to test in streams. This was surprising since many sampling based techniques can be extended to data streams. For example, Bar-Yossef et al [10] showed that non-adaptive sampling can be easily simulated in an aggregate (all occurrences of item $i$ are grouped together) streaming model with a small blowup in space. The aggregation assumption can be removed with an extra pass.

We show that in fact these (variants of permutations) are the only hard functions. Specifically, we show that any property testing algorithm for a permutation invariant (also known as symmetric) function, in the *combined* oracle model can be simulated by a single pass data stream algorithm that assumes a random permutation of the input. The random permutation assumption can be removed using an extra pass to give a two-pass simulation in the regular streaming model. The simulation builds upon the reductions used by Bar-Yossef et al [8, 6, 7] in deriving strong lower bounds for sampling. However we use the reductions for upper bounds.

In a natural sense, if we exclude permutation dependent functions, stream testing in the random permutation model subsumes combined oracle property testing, and it is the testing of entropy that reveals this difference between the models.

## 2 Definitions

**Definition 1.** *Let $p$ and $q$ be two discrete probability distributions defined on base $[n]$. The $f$-divergence [18] between $p$ and $q$ is defined as $D_f(p,q) = \sum p_i f(q_i/p_i)$ for some function $f$ (convex, $f(1) = 0$). Many commonly used distance measures are $f$-divergences, including the $\ell_1$ distance, the* Hellinger distance[3] *Hellinger$(p,q) = \sum_i (\sqrt{p_i} - \sqrt{q_i})^2$, the Jensen-Shannon distance $JS(p,q) = \sum_i p_i \ln \frac{2p_i}{p_i+q_i} + q_i \ln \frac{2q_i}{p_i+q_i}$ and the* Triangle distance $\Delta(p,q) = \sum_i \frac{(p_i-q_i)^2}{p_i+q_i}$.

**Definition 2.** *The entropy of a distribution is defined as $H(p) = \sum_i p_i \log \frac{1}{p_i}$. (All logs are base 2.)*

Accurately estimating entropy is an interesting problem in its own right as well from the context of the statistical distances. Throughout we will make use of the following lemma.

**Lemma 3 (Concentration of Independent Random Variables [25]).** *Let $\{X_t\}_{1 \leq t \leq m}$ be independently distributed random variables with (continuous) range $[0, u]$. Let $X = \sum_{1 \leq t \leq m} X_t$. Then for $\gamma > 0$, $\mathbb{P}(|X - \mathbb{E}(X)| \geq \gamma \mathbb{E}(X)) \leq 2 \exp(\frac{-\gamma^2 \mathbb{E}(X_t) m}{3u})$.*

## 3 Property Testing

### 3.1 Oracle Models for Property Testing of Distributions

Two main oracle models have been used in the property testing literature for testing properties of distributions. These are the *generative* and *evaluative* models introduced by Kearns et al [29]. The black-box or generative model of a distribution permits only one operation: taking a sample from the distribution. In other words, given a distribution $p = \{p_1, \ldots p_n\}$, `sample(p)` returns $i$ with probability $p_i$. In the evaluative model, a `probe` operation is permitted. `probe(p, i)` returns $p_i$. A natural third model, the *combined* model was introduced by Batu et al [11]. In this model both the `sample` and `probe` operations are permissible. In all three models, the complexity of an algorithm is measured by the number of operations.

---
[3]Note that the Hellinger distance is sometimes defined as the square-root of the above quantity.

### 3.2 Testing Jensen-Shannon, Hellinger and Triangle Divergences (Generative Oracle)

In this section we consider property testing in the generative model for various information theoretic distances. We will present the results for the Triangle Divergence $\Delta$. However, the Jensen-Shannon and Hellinger divergences are constant factor related to the Triangle divergence as follows:

$$\text{Hellinger}(p, q)/2 \leq \Delta(p, q)/2 \leq JS(p, q) \leq \ln 2 \Delta(p, q) \leq 2 \ln 2 \text{Hellinger}(p, q)$$

(Parts of equation 3.2 are proved in [37].) Therefore the results presented here naturally imply analogous results for them as well. Our algorithm is similar to that in [12], and is presented in Figure 1. It relies on an $\ell_2$ tester given in [12]. Central to the analysis are the following inequalities.

$$\frac{\ell_2^2(p,q)}{\max(\ell_\infty(p) + \ell_\infty(q))} \leq \Delta(p,q) \leq \ell_1(p,q) \leq \sqrt{n}\ell_2(p,q) \tag{1}$$

**Lemma 4 ($\ell_2$ Testing [12]).** *There exists an algorithm that given distributions $p, q$ draws $s = O(\log \frac{1}{\delta}(b^2 + \epsilon^2\sqrt{b})/\epsilon^4)$ samples (where $b = \max_i(p_i, q_i)$) and if $\ell_2(p, q) \leq \epsilon/2$, the algorithms passes with probability at least $1 - \delta$, but if $\ell_2(p, q) \geq \epsilon$ the algorithm passes with probability less than $\delta$.*

> **Algorithm $\Delta$-*Test***
> 1. Draw $m$ samples from $p$ and from $q$
> 2. $n_i^p = \#$ times element $i$ appears and $\tilde{p}_i = n_i^p/m$
> 3. $n_i^q = \#$ times element $i$ appears and $\tilde{q}_i = n_i^q/m$
> 4. Let $S = \{i : \max\{n_i^p, n_i^q\} \geq mn^{-\alpha}\}$
> 5. **return** "Fail" if $\sum_{i \in S} \frac{(\tilde{p}_i - \tilde{q}_i)^2}{\tilde{p}_i + \tilde{q}_i} > \epsilon/10$
> 6. Define $p'$ (and $q'$ analogously) as follows:
> 7. $i \leftarrow$ `sample(p)`
> 8. if $i \notin S$, output $i$ else output $j$ uniformly chosen from $[n]$
> 9. **return** $\ell_2$-Tester ran on $p', q'$ and $\frac{\epsilon}{\sqrt{n}}$

Figure 1: $\Delta$-Testing in the Generative Model

The first lemma follows from Chernoff bounds. The rest of the proofs are in Appendix A.

**Lemma 5.** *We say an estimate is* heavy *if it is greater than $1/n^\alpha$. Then, with $m = O(\log \frac{1}{\delta} \frac{n^\alpha \log n}{\gamma^2})$ samples, with probability $1 - \delta/2$, for all heavy estimate $\tilde{p}_i$, $\tilde{p}_i$ is at most $p_i \gamma/100$ from $p_i$. Furthermore, if at least one of $\tilde{p}_i$ or $\tilde{q}_i$ is heavy then we can estimate $\frac{(p_i - q_i)^2}{p_i + q_i}$ up to $\pm \gamma \frac{\max\{p_i, q_i\}}{10}$.*

**Lemma 6 ($\Delta$ Testing).** *For two distributions $P$ and $Q$, then there exists an algorithm drawing*

$$s = O(\log \frac{1}{\delta} \max\{\frac{n^\alpha \log n}{\epsilon^2}, (n^{-2\alpha+2} + \epsilon^2 n^{1-\alpha/2})/\epsilon^4\})$$

*samples such that if $\Delta(P, Q) \leq \epsilon^2/n^{1-\alpha}$, the algorithms passes with probability at least $1 - \delta$, but it $\Delta(P, Q) \geq \epsilon$ the algorithm passes with probability less than $\delta$.*

Observe that setting $\alpha = 2/3$ yields an algorithm with sample complexity $\tilde{O}(n^{2/3}/\epsilon^4)$.

### 3.3 Testing all Symmetric Bounded $f$-Divergences (Combined Oracle)

In this section we consider property testing in the combined oracle model for all symmetric bounded $f$-divergences. Recall that a convex function $f$ defines a divergence $D_f(p,q) = \sum_i p_i f(\frac{q_i}{p_i})$; this encodes the permutation invariance. We are interested in symmetric (over $p,q$) divergences, i.e., $D_f(p,q) = D_f(q,p)$. We define a divergence to be *bounded*[4] if $\max\{f(u), uf(\frac{1}{u})\} = \tau < \infty$. JS, Hellinger, $\Delta, \ell_1$ all are bounded and satisfy $\tau = 2$. We show an interesting decomposition property of symmetric $f$ divergences. Define a *conjugate* $f^*(u) = uf(\frac{1}{u})$. It can be verified that JS, Hellinger, $\Delta, \ell_1$ are self conjugates, i.e., $f^*(u) = uf(\frac{1}{u})$. One useful characterization of symmetric $f$-divergence is the following:

**Lemma 7 ([31]).** $D_f(p,q) = D_f(q,p)$ iff $f^*(u) = f(u) - c(u-1)$ for some constant $c$.

Therefore using $f$ is the same as using $\frac{f(u)+f^*(u)-2c(u-1)}{2}$, but since $\sum_i p_i c(\frac{q_i}{p_i} - 1) = c(\sum_i q_i - \sum_i p_i) = c(1-1) = 0$, we may as well use $g(u) = \frac{f(u)+f^*(u)}{2}$. We now claim the following:

**Lemma 8.** *Any symmetric divergence $f$, can be expressed as* $D_f(p,q) = \sum_{i:p_i > q_i} p_i g(x) + \sum_{i:q_i > p_i} q_i g(1/x)$
*where $g(x) = \frac{1}{2}(f(x) + xf(\frac{1}{x}))$ and $x = \frac{q_i}{p_i}$. Further, if $f$ is bounded then $g(x) \leq \tau$ if $x \in [0,1]$.*

Although the above appears simple, it actually allows us to *break* the divergence into small, positive components. This allows us to use sharp concentration bounds.

---

**Algorithm** *Combined Oracle Distance Testing*
1.     $E \leftarrow 0$ **for** $t = 1$ to $m$:
2.        **do** $i \leftarrow \texttt{sample}(p)$ and $x = \texttt{probe}(q,i)/\texttt{probe}(p,i)$
3.           If $x > 1$ then $a \leftarrow g(x)$ else $a \leftarrow 0$
4.           $j \leftarrow \texttt{sample}(q)$ and $x = \texttt{probe}(q,j)/\texttt{probe}(p,j)$
5.           If $x < 1$ then $b \leftarrow g(1/x)$ else $b \leftarrow 0$
6.           $E \leftarrow (a+b)/2\tau + E$
7.     Estimate the distance as $2\tau E/m$

---

Figure 2: Combined Oracle Distance Testing

**Theorem 9.** *For a bounded symmetric $f$-divergence $D_f$ in the combined oracle model we can estimate $D_f(p,q)$ upto a factor $(1+\epsilon)$ in $O(\tau/(\epsilon^2 D_f(p,q)))$ time. ($\tau = O(1)$ for $\ell_1$, Hellinger, $\Delta$, and JS.)*

*Proof.* Consider the value $\frac{a+b}{2u}$ added to $E$ in each iteration. This is a random variable with range $[0,1]$ and mean $\frac{D_f(p,q)}{2\tau}$. Hence by Lemma 3, $\mathbb{P}\left(|E - m\frac{D_f(p,q)}{2\tau}| < \epsilon m \frac{D_f(p,q)}{2\tau}\right) \leq 2\exp(-\epsilon^2 D_f(p,q)m/6\tau)$. Therefore with $O(\tau/(\epsilon^2 D_f(p,q)))$ samples/probes the probability that we do not estimate $D_f(p,q)$ as required can be made arbitrarily small. □

Note that although $\ell_2$ is not an $f$-divergence, setting $a = p_i(1-q_i/p_i)^2$ and $b = q_i(p_i/q_i - 1)^2$ in the above we can estimate $\ell_2^2(p,q)$ in $O(1/(\epsilon^2 \ell_2^2(p,q)))$ time. It is worth mentioning that the above results can be rephrased as a $O(1/\epsilon)$ algorithm if we are interested in distinguishing between the cases where the distance is greater than $\epsilon$ or less than $\epsilon/2$.

We now prove a corresponding lower bound that shows that our algorithm is tight. Note that while it is relatively simple to see that there exists two distributions that are indistinguishable with less than $o(1/\ell_1)$ oracle calls, it requires some work to also show a lower bound with a dependence on $\epsilon$. Further note that the below proof also gives analogous results for JS, Hellinger and $\Delta$. (This is because for any pair of probability distributions $p$ and $q$ such that $p_i \neq q_i \Rightarrow \min\{p_i, q_i\} = 0$ then $\ell_1(p,q) = \Delta(p,q) = \mathrm{JS}(p,q) = \mathrm{Hellinger}(p,q)$.)

---

[4]Since $f$ is typically monotone, i.e., if the likelihood ratio is closer to 1, then the distance decreases, and so boundedness reduces to $\lim_{u \to 0} f(u)$ and $\lim_{u \to \infty} \frac{f(u)}{u}$ exist and is at most $\tau < \infty$. Note that from continuity $\lim_{u \to 1} f(u) = f(1) = 0$.

**Theorem 10 ($\ell_1$ Lower Bound in the Combined Oracle Model).** *Any approximation up to a $1 + \epsilon$ factor of the $\ell_1$ distance between two distributions in the combined oracle model requires $\Omega(\frac{1}{\epsilon^2 \ell_1})$ time.*

*Proof.* Let $p$ and $q^r$ be the distributions on $[n]$ described by the following two probability vectors:

$$p = (1 - 3a/2, \overbrace{3a\epsilon/2k, \ldots, 3a\epsilon/2k}^{k/\epsilon}, 0, \ldots, 0) \text{ and } q^r = (1 - 3a/2, \overbrace{0, \ldots 0}^{r}, \overbrace{3a\epsilon/2k, \ldots, 3a\epsilon/2k}^{k/\epsilon}, 0, \ldots, 0)$$

Then $\ell_1(p, q^{k/3\epsilon}) = a$ and $\ell_1(p, q^{k/3\epsilon+k}) = a(1 + 3\epsilon)$. Hence to $1 + \epsilon$ approximate the distance between $p$ and $q^r$ we need to distinguish between the cases when $r = k/3\epsilon (=: r_1)$ and $r = k/3\epsilon + k (=: r_2)$. Consider the distributions $p'$ and $q^{r\prime}$ formed by arbitrarily permuting the base sets of the $p$ and $q^r$. Note that the $\ell_1$ distance remains the same. We will show that, without knowledge of the permutation, it is impossible to estimate this distance with $o(1/(\epsilon^2 a))$ oracle calls. We reason this by first disregarding the value of any "blind probes", ie. a probe $\mathtt{probe}(p', i)$ or $\mathtt{probe}(q', i)$ for any $i$ that has not been returned as a sample. This is the case because, by choosing $n \gg k/(a\epsilon^2)$ we ensure that, with arbitrarily high probability, for any $o(1/(\epsilon^2 a))$ set of $i$'s chosen from any $n - o(1/(a\epsilon^2))$ sized subset of $[n]$, $p_i' = q_i^{r\prime} = 0$. This is the case for both $r_1$ and $r_2$. Let $I = \{i : p_i \text{ or } q_i^r = 3a\epsilon/2k\}$ and $I_1 = \{i \in I : p_i \neq q_i^r\}$. Therefore determining whether $r = r_1$ or $r_2$ is equivalent to determining whether $|I_1|/|I| = 1/2$ or $1/2 + \frac{9\epsilon}{8+6\epsilon}$. We may assume that every time an algorithm sees $i$ returned by $\mathtt{sample}(p)$ or $\mathtt{sample}(q)$, it learns the exact values of $p_i$ and $q_i$ for free. Furthermore, by making $k$ large ($k = \omega(1/\epsilon^3)$ suffices) we can ensure that no two $\mathtt{sample}$ oracle calls will ever return the same $i \in I$ (with high probability.) Hence distinguishing between $|I_1|/|I| = 1/2$ and $1/2 + \frac{9\epsilon}{8+6\epsilon}$ is analogous to distinguishing between a fair coin and a $\frac{9\epsilon}{8+6\epsilon} = \Theta(\epsilon)$ biased coin. It is well known that the latter requires $\Omega(1/\epsilon^2)$ samples. Unfortunately only $O(1/a)$ samples return an $i \in I$ since with probability $1 - 3a/2$ we output an $i \notin I$ when sampling either $p$ or $q$. The bound follows. □

### 3.4 Testing Entropy (Combined Oracle)

In this subsection we show that a simple algorithm achieves the optimal bounds for estimating the entropy in the combined oracle model of property testing. Note that this simple algorithm improves upon the previous upper bound of Batu et al [11] by a factor of $\log n/H$ where $H$ is the entropy of the distribution. The authors of [11] showed that their algorithms were tight for $H = \Omega(\log n)$; *we show that the upper and lower bounds match for arbitrary $H$*. The algorithm is presented in Figure 3. It is structurally similar to the algorithm given in [11] but uses a cutoff that will allow for a much tighter analysis via Chernoff bounds.

**Algorithm** *Combined Oracle Entropy Testing*
1. $E \leftarrow 0$ **for** $t = 1$ to $m$:
2.    **do** $i \leftarrow \mathtt{sample}(p)$
3.      $p_i \leftarrow \mathtt{probe}(p, i)$
4.      if $p_i \geq \frac{1}{n^3}$ then $a \leftarrow \frac{\log 1/p_i}{3 \log n}$ else $a \leftarrow 0$
5.      $E \leftarrow a + E$
6. Estimate the entropy as $3E \log n/m$

Figure 3: Combined Oracle Entropy Testing

The next lemma estimates the contribution of the unseen elements and that leads to the main theorem about estimating entropy in the combined oracle model.

**Lemma 11.** *Consider any set subset $S$ of $[n]$, then $\sum_{i \in S} p_i \log 1/p_i \leq \sum_{i \in S} p_i \log \frac{|S|}{\sum_{i \in S} p_i}$. In particular, if $\sum_{i \in S} p_i \leq \frac{\epsilon H(p)}{\log n - \log \epsilon H(p)}$ then $\sum_{i \in S} p_i \log 1/p_i \leq \epsilon H(p)$.*

**Theorem 12.** *In the combined oracle model we can $1 + \epsilon$ estimate the entropy $H(p)$ of a distribution using $O(\frac{\log n}{\epsilon^2 H(p)})$ where $H$ is the entropy of the distribution.*

*Proof.* We restrict our attention to the case when $H(p) > 1/n$ and $\epsilon > 1/\sqrt{n}$ since otherwise we can trivially find the entropy exactly in $O(1/\epsilon^2 H(p))$ time by simply probing each of the $n$ $p_i$'s. Consider the value $a$ added to $E$ in each iteration. This is a random variable with range $[0, 1]$ since $p_i \geq 1/n^3$ guarantees that $\frac{\log 1/p_i}{3 \log n} \leq 1$. Now, the combined mass of all $p_i$ such that $p_i < 1/n^3$ is at most $1/n^2$. Hence since $1/n^2 \leq \frac{\epsilon/2 H(p)}{\log n - \log \epsilon/2 H(p)}$ by lemma 11 the maximum contribution to the entropy from such $i$ is at most $\epsilon H(p)$. Hence the expected value of $a$ between $(1 - \epsilon/2)H(p)/3 \log n$ and $H(p)/3 \log n$ and therefore, if we can $1 + \epsilon/2$ approximate $\mathbb{E}(a)$ then we are done. We use the probability concentration lemma 3 to get that $\mathbb{P}(|E - m\mathbb{E}(a)| < (\epsilon/2)m\mathbb{E}(a)) \leq 2e^{-(\epsilon/2)^2 m \mathbb{E}(a)/3}$. Therefore with $O(1/(\epsilon^2 \mathbb{E}(a))) = O(\log n/\epsilon^2 H(p))$ samples/probes the probability that we don't $1 + \epsilon/2$ approximate $\mathbb{E}(a)$ estimate $H(p)$ can be made arbitrarily small. □

## 4 Data Streams

### 4.1 The Data Streams Model

The data stream model characterizes small space algorithms that can access the read-only input in order. The algorithm makes passes over the input; any item not explicitly stored is inaccessible to the algorithm in the same pass. In many cases the number of passes is limited to one. The crucial aspect of a (regular) data stream algorithm is that the algorithm is required to produce a correct output for an arbitrary permutation of the input stream.

As mentioned in the introduction, *a random stream algorithm* is a data stream algorithm that reads a *randomly permuted* input from its read-only input tape. Alternate definitions are possible, but this definition dates back to Munro and Paterson [33] and we will restrict ourselves to this definition. (It was also appear in [20].) All other features are the same as a general stream algorithm. As usual, the complexity of the algorithm is measured primarily in terms of the amount of space used on the work tape (for which the algorithm has random read-write access).

*Modeling stream distributions:* There are two ways in which a data stream can be considered to define a probability distribution $p$. These are the *update data stream model* and the *aggregate data stream model*. Firstly we discuss the *update data stream model*. We are given a base domain $[1, \ldots, n]$ over integers and a function $f_p()$ is specified as $\langle p, i, + \rangle$ which corresponds to $f_p(i) \leftarrow f_p(i) + 1$. This is the model used by Alon, Matias, and Szegedy in [2]. The model naturally captures $f_p(i) \leftarrow f_p(i) + \Delta_i$, however we do not consider $f_p(i) \leftarrow f_p(i) - 1$ (deletions) since the negative term does not correspond to any operation over distributions.

An alternate model is the *aggregate model* where the input is $\langle p, i, f_p(i) \rangle$. This is the model used by Feigenbaum et al [23] for $\ell_1$ differences[5]. In this model, computing the entropy is trivial and the Hellinger distance reduces to computing the $\ell_2^2$ norm[6]. Note that this implies that we can compute the Jensen-Shannon

---
[5] Later generalized by Indyk [26] to the update model.

[6] Suppose we are given streams $X$ and $Y$ representing distributions $P$ and $Q$ respectively. Let $f_p(i), f_q(i)$ be the number of occurrence of item $i$ in the two streams and let $m_x = \sum_i f_p(i), m_y = \sum_i f_q(i)$ be the lengths of the streams respectively. The (squared) Hellinger distance is the $\ell_2^2$ norm of the difference of vectors $\mathbf{u}, v$ where $u_i = \sqrt{f_p(i)/m_x}, v_i = \sqrt{f_q(i)/m_y}$. This difference can be easily computed by sketching (a la Johnson Lindenstrauss [28], see [26]), maintaining the product $A\mathbf{u}$ (and scaling by $\sqrt{m_x}$ at end of input).

Divergence, the Triangle-divergence upto a constant factor as well. Obtaining a PTAS for them remains an interesting open question even in this simpler model. However, the aggregate model is restrictive for distributions, since the aggregation loses the "distribution" aspect. We will focus on the update model, and, as we argued above, will only consider insertions.

*Random Streams and two distributions:* When we are computing a function of two distributions $p$ and $q$ we also have a function $f_q()$ specified by data items $\langle q, i, +\rangle$. Note that, for a random stream algorithm, we consider the random permutation to be over $\langle q, i, +\rangle$ and $\langle p, i, +\rangle$ together. We will assume that $\sum_i f_p(i)$ and $\sum_i f_q(i)$ are within a constant factor of each other.

## 4.2 Simulating the Combined Oracle Models

In the next section we will discuss how algorithms that make (combined) oracle calls may be simulated in the various streaming models. In particular this leads to the following theorem.

**Theorem 13.** *In two passes of a regular stream, there exist algorithms that,*

1. $(1+\epsilon)$ approximate the entropy $H$ using $\tilde{O}(\frac{1}{\epsilon^2 H})$ space.

2. $(1+\epsilon)$ approximate a $\tau$ bounded symmetric $f$-Divergence $D_f$ using $\tilde{O}(\frac{\tau}{\epsilon^2 D_f})$ space.

*If the stream is randomly ordered then one pass suffices in each case.*

Unfortunately, it is sometimes unrealistic to assume more than a single pass over the data. Hence we now concentrate on single pass algorithms. In what follows we present a single pass, polylog space, asymptotic constant factor approximation in a single pass. We then briefly discuss a single pass algorithm that uses $\tilde{O}(n^\alpha)$ space and achieves a $(1+\epsilon)\frac{1}{\alpha}$ approximation. Finally we present one pass algorithm for the random streaming model whose memory needs are not in terms of $1/H$

## 4.3 An Asymptotic Approximation Scheme for Estimating Entropy in Regular Streams

To construct our algorithm we will use algorithms for approximating the $F_0$. There has been a long history of papers for computing the frequency moments of streams. We focus our attention to the best known $(\epsilon, \delta)$ approximation algorithm of Bar-Yossef et al [9], where $F_0$ is approximated upto a factor $(1+\epsilon)$ with probability $1 - \delta$. Their result shows that the $(\epsilon, \delta)$ approximation can be performed in $O((\frac{1}{\epsilon^2} \log \log n + \log n) \log \frac{1}{\delta})$ space. We will only focus on the fact that the space bounds are poly-logarithmic. The basic intuition of the algorithm is similar to the sublinear time minimum spanning tree algorithm of Chazelle et al [15] and the streaming geometry algorithms by Indyk [27]. The idea is to count objects at various resolutions.

Our algorithm works by randomly generating conceptual sub-streams from the data stream. Each sub-stream has a associated level $j$ and we will perform the random generation of a sub-stream of level $j$ in such a way that we only expect elements $i$ with $p_i \geq 2^{-j}$ to appear in the stream. We will feed each sub-stream into an algorithm for estimating the number of distinct elements in the sub-stream. Summing up the these estimates (appropriately scaled) will give our estimate. The net result will be an asymptotic approximation scheme of factor $\frac{e}{e-1}$, i.e., for $H$ sufficiently large (but constant)[7]. The algorithm is presented in Figure 4.

The centerpiece of the algorithm are the following two Lemmas which are proved in Appendix A. The bounds follows from the accuracy of our guess $\tilde{m}$ for $m$. Let $\chi_{ij}$ be the event that item $i$ showed up in level $j$. Note that $\chi_{ij}$ are independent.

**Lemma 14.** *If $m = \Omega(\frac{1}{\epsilon^2}), p_i 2^j \leq 1$ and $t \geq (1+\epsilon)^2$ then $(1 - \frac{1}{e})\frac{2^j p_i}{t} \leq Pr[\chi_{ij} = 1] \leq (1+\epsilon)^2 \frac{p_i 2^j}{t}$.*

---
[7]This is in the same spirit as the approximation algorithms for bin-packing, checking for packing in 2 bins is NP-HARD, but we have a PTAS as the number of bins is large.

> **Algorithm** *Entropy-Estimation*
> 1. Guess $\tilde{m} = 1, (1+\epsilon), \ldots (1+\epsilon)^i$ where $\tilde{m}$ is roughly the number of insertions.
> 2. Initially $m = 0$, we will keep track of $m$.
> 3. **for** each guess of $\tilde{m}$
> 4.     **do for** each stream item $\langle i, + \rangle$
> 5.         **do** For $j = 1, \ldots k$, where $k \approx \log \frac{n}{\epsilon}$, toss a coin with probability $\frac{2^j}{\tilde{m}t}$ (independent across $j$) and include $i$ in sub-stream $j$. $t \sim (1+\epsilon)^2$ is almost 1.
> 6.         Maintain appropriate structure for computing $F_0$ in each of these $k$ sub-streams.
> 7.     Choose the set of structures corresponding to the guess $(1+\epsilon)\tilde{m} > m \geq \tilde{m}$.
> 8.     For this $\tilde{m}$, for each $j$ estimate the number of distinct items $f_j = F_0(j)$.
> 9.     Output $\sum_j f_j / 2^j$.

Figure 4: The $F_0$ Algorithm for Estimating Entropy in a Data Stream

**Lemma 15.** $\frac{1}{t}(1 - \frac{1}{e})(H-1) \leq \mathbf{E}\left[\sum_j \frac{f_j}{2^j}\right] = \mathbf{E}\left[\sum_j \frac{\sum_i Pr[\chi_{ij}=1]}{2^j}\right] \leq \frac{(1+\epsilon)^2}{t}H + 2$ *where $H$ is the entropy.*

We assume that the length of the stream is $m \gg n/\epsilon$ but that $m$ is polynomial in $n$. Now observe that $\sum_j f_j/2^j$ is a sum of many (bounded) Bernoulli variables. We can easily apply Chernoff and show with probability $O(\exp(-\epsilon_c^2 H))$ the sum $\sum_j f_j/2^j \in \left[(1-\frac{1}{e})(H-1)(1-\epsilon_c), (\frac{1}{t}(1+\epsilon)^2 H + 2)(1+\epsilon_c)\right]$. We need to repeat the above for $O(\log n/\epsilon_c H)$ times for high probability. Now, we lose a further $1 \pm \epsilon_0$ factor in the $F_0$ estimation. Overall, we maintain $O(\log n/\epsilon)$ different $F_0$ estimation structure, each of which takes space $O((1/\epsilon_0^2) \log \log n + \log n) \log n$ and succeed with high probability. Finally there are $\log n/\epsilon$ guesses for $\tilde{m}$. The overall space bound is $O((\log^3 n/(\epsilon \epsilon_c H))(\frac{1}{\epsilon_0^2} \log \log n + \log n))$.

**Theorem 16.** *We have a polylog space asymptotic $\frac{e}{e-1} + \epsilon$ approximation for the entropy.*

A natural question is if the above analysis is tight. It is possible to slightly improve the bounds in Lemma 15 but the problem is that there will always be constant shift between the entropy and our estimate. There is a natural bias in the estimation entropy and this particular method alone is unlikely to yield better results.

### 4.4 A True PTAS (at a Price) for Estimating Entropy in Regular Streams

A final natural question remains, namely, can we achieve a PTAS for entropy in the regular data streams model? The answer is yes, but, it comes with a price. The space bound for $\approx \frac{1}{\alpha}$ approximation increases over the space bound of the combined oracle model by a factor approximately $n^\alpha/H$. The algorithm and its analysis is presented in Appendix B. This algorithm is roughly analogous to the algorithm presented in Sec. 4.5. The difference lies in the way it divides the elements into "large" and "small" classes. It proceeds in two steps (i) Finds the elements with large $f(i)$ and estimate them and (ii) Uses a worst case bound for small $f(i)$. The first step is achieved by a careful sampling technique reminiscent of the online facility location algorithm of Meyerson [32] in the context of stream clustering. There are also some similarities with the count/count-min sketches of [14, 16].

**Theorem 17.** *We can estimate the entropy $H$ upto a factor $\frac{1}{\alpha}(1+\epsilon')$ using space $O(\frac{n^\alpha \log n}{f(\epsilon', H)})$ where $f(\epsilon', H)$ is the value of $\epsilon$ satisfying, $\epsilon' = \frac{\alpha(\epsilon \log(n/\epsilon) + n^{-\alpha})}{H} + \frac{\log 1/\epsilon}{\log n}$.*

### 4.5 A PTAS in the Random Streams Model

Emulating the combined oracle algorithm is only inefficient when the entropy is very small. But when the entropy is small, it is easy to see that there must be one element with probability mass $\approx 1$. The high level idea of our algorithm is to keep track the exact counts of the elements of a certain set $A$; and establish that the projection of the distribution (rescaled to 1) to the complement of $A$ has large entropy. On this projection we run the simulation of the combined oracle algorithm. However, we have several problems; (a) when we discover that entropy is large we have already a few elements from the projection – we have a dependence, (b) the projected distribution may again have one element with mass $\approx 1$, and (c) we will not know the exact probability mass of any set till the end of input. But,

**Proposition 18.** *In the random stream model if we consider the first $s$ elements which do not belong to $A$, we have a random sample (without replacement) from the distribution projected to $[n] - A$.*

Using Proposition 18, we can show that either estimates are sufficient for (b) and (c), or the stream has few distinct elements, which we count exactly.[8] At the end of input we will know the exact probability mass of $A$ and rescale/shift to get the contribution of the projection to the original distribution. The proof of the next lemma is in Appendix A.

**Lemma 19.** *We can eliminate the dependence between $MS$ and the simulated property testing algorithm.*

**The algorithm:** At any point of time we are maintaining a set of items $A$. For each $i \in A$ we are counting the number of occurrences of $i$. We keep seeing elements in the stream until we accumulate a multi-set $MS$ of size $c_1 \log n$ of elements that do not belong to $A$ and appear earliest in the stream for some large constant $c_1$. We check if any $i' \in MS$ occurs at least $\frac{c_1}{2} \log n$ times in $MS$.

- If there is no such $i'$ (projection has large entropy), we proceed to simulate the $t$ queries with the correction due to $M$ as described above (maintaining counts of the elements in $A$ as well).

- If there is any such $i'$, let $A \to A \cup \text{distinct}(MS)$. We operate on the rest of the stream with this new $A$. Observe that the invariant of maintaining the count of elements in $A$ can be maintained.

We also store the last $O(\epsilon^{-2} \log n)$ elements in the stream (also a multi-set) seen at all time.

Note that we can assume that the remaining stream is much larger than $MS$ – we maintain the last $O(\epsilon^{-2} \log n)$ elements and would discover that the remaining stream is small within our alloted space. But in that event we would have an exact description of the distribution of the items in the stream. The next lemma follows from the assumption, Proposition 18 and the Chernoff bound.

**Lemma 20.** *Assuming that the remaining stream is much larger than $MS$, if there is an item in the projection with mass $\geq \frac{1}{4}$, then it occurs at least $\frac{1}{2}$ fraction in $MS$ with high probability polynomially close to $1$.*

The theorem follows from the fact that either we take out probability mass quickly, or shift to the simulation phase. The proof is in Sec. A.

**Theorem 21.** *We can compute a $(1+\epsilon)$-approximation for the entropy in a random stream model using $O((\log n + \epsilon^2) \log n)$ space.*

We can tighten the above by noticing that if the number of distinct elements in $MS$ is $r+1$ then entropy is $\Omega(r/\log n)$ and we can shift to the second phase. We omit the discussion.

---

[8]We can view the setting as a "robust distribution" as in Gilbert et al [24]

## 5 Connecting Oracle Models and Streaming Models

We direct the reader to [6] for a detailed treatment of the relative computational power of the data stream, sketch (see [6] for a definition) and generative sampling models. Here we restrict ourselves to comparing the combined oracle model with the streaming model.

We say that a function $f(p) = f(p_1, p_2, \ldots p_n)$ is *symmetric* (or *permutation-invariant*), if $f$ remains unchanged when its arguments are permuted. (Symmetry is a desirable and often-assumed property of functions on distributions, and is a special case of general invariance under coordinate reparametrizations [13].) We will show that we can always express the algorithm in a canonical form where the algorithm *first samples* and then probes the samples along with a few other elements. The idea would be to view the original algorithm, after the sampling stages and probing of the samples, as a randomized decision tree that we rewrite to an oblivious decision tree along the lines of Bar-Yossef et al [10, 6], and simulating this new decision tree in the random stream model. We start with the necessary definitions.

**Definition 22.** *A randomized decision tree that computes a function $f(x)$ is defined (as usual) as a decision tree having three types of nodes; a* query *node that asks for the value of an input parameter and maps the resulting value (and the history of all queries upto that point) to a choice of child node to visit, a random choice node, where the child node is chosen at random, and output nodes, where an answer expressed as a function of all queries thus far is returned.*

**Definition 23.** *An* oblivious decision tree *is one where the queries are made independent of the input, or the random choices in the algorithm. Formally, suppose we have a tree $T$ with worst-case query complexity $u$. Then an* I-relabeling *of $T$ by $I = \{i_1, \ldots i_u\}$ relabels all query nodes of depth $j$ by the query to $i_j$, yielding the tree $T^I$. An oblivious decision tree is then a pair $T, \Delta_u$, where $T$ is a decision tree with worst-case complexity $q$ and $\Delta_u$ is a distribution on $[n]^u$. A computation on an oblivious decision tree consists of two steps: (1) sample $u$ elements $I$ from $\Delta_q$, (2) Relabel $T$ to $T^I$ and run it on input $x$.*

The first lemma shows how any combined oracle tester can be transformed (with only a slight blow-up in complexity) to one of a canonical form. The proof is is deferred to Appendix A.

**Lemma 24 (Canonical Form Algorithm).** *Let $\mathcal{A}$ be a combined oracle property testing algorithm that $1 + \gamma$ estimates a symmetric function $f(p)$ using (worst-case) $t$ oracle calls and probability of error $\delta$. Then there exists a canonical algorithm $\mathcal{A}'$ that uses (worst case) $O(t)$ oracle calls with equal performance.*

We are now ready to prove the main structural result of this section. The proof is deferred to Appendix A. The main idea for simulating in two pass regular stream model is to sample in the first pass and then do exact counting in the second pass. For the random order stream result we are able to do both the sampling and exact counting in the same pass by using, roughly speaking, the prefix of the random order stream as a source for sample oracle queries.

**Theorem 25.** *Let $p$ be the probability distribution described by an update data stream. Let $\mathcal{A}$ be a combined oracle property testing algorithm that makes at most $t$ oracle calls to $1 + \gamma$ estimate a symmetric function $f(p)$ with probability of error at most $\delta$. Then there exist a single pass random stream algorithm and a two pass regular stream algorithm that use $O(t \log m \log n)$ space with equal performance.*

The proof can be generalized to the case when we are computing a function of two distributions $f(p, q)$, e.g., a distance between two distributions. In this case we consider $f$ as a function over $n$ tuples, ie. $f(p, q) = f((p_1, q_1), (p_2, q_2), \ldots (p_n, q_n))$. $f$ is symmetric is it is invariant of permutations of the $n$-tuples. The only important caveat is that we need $\sum_i f_p(i) = \Theta(\sum_i f_q(i))$ such that, with high probability, there are $t$ elements of the form $\langle p, i, +\rangle$ (for some $i$) and $t$ elements of the form $\langle q, i, +\rangle$ (for some $i$) in the first $O(t)$ data items.

## A  Miscellaneous Proofs

*Proof of Lemma 5.* Wlog. let $\tilde{p}_i \geq \tilde{q}_i$. Let $\tilde{p}_i = (1+\gamma_1)p_i$, $\tilde{q}_i = (1+\gamma_2)q_i$ and $\gamma' = \gamma/100$.

$$\begin{aligned}
& \left|\frac{(\tilde{p}_i - \tilde{q}_i)^2}{\tilde{p}_i + \tilde{q}_i} - \frac{(p_i - q_i)^2}{p_i + q_i}\right| \\
\leq\ & \left|\frac{(\tilde{p}_i - \tilde{q}_i)^2 - (p_i - q_i)^2}{p_i + q_i}\right| + \left|\left(\frac{p_i + q_i}{\tilde{p}_i + \tilde{q}_i} - 1\right)\frac{(\tilde{p}_i - \tilde{q}_i)^2}{p_i + q_i}\right| \\
\leq\ & \left|\frac{(\tilde{p}_i - \tilde{q}_i)^2 - (p_i - q_i)^2}{p_i + q_i}\right| + |\gamma'(1+\gamma')|p_i \\
\leq\ & \left(\left|(2\gamma_2 + \gamma_2^2)\left(\frac{q_i}{p_i}\right)^2\right| + \left|2\left(\frac{q_i}{p_i}\right)(\gamma_1 + \gamma_2 + \gamma_1\gamma_2)\right| + |(2\gamma_1 + \gamma_1^2)| + |\gamma'(1+\gamma')|\right)p_i
\end{aligned}$$

Wlog, $p_i > q_i$. Now if $p_i, q_i \geq n^{-\alpha}$ we know that $|\gamma_1|, |\gamma_2| \leq \gamma'$ can thus we are bounded above by $p_i\gamma/10$. The problem is that is $q_i \ll n^{-\alpha}$ then $|\gamma_2|$ could be very large. However in this case $q_i/p_i$ will be very small so we can still bound the above well. Formally, let $q_i = \lambda_1 n^{-\alpha}$ and $n^{-\alpha} = \lambda_2 p_i$. Note that $0 \leq \lambda_1 < 1$ and $0 < \lambda_2 < (1+\gamma)$. Now, note that if $m$ samples is sufficient to $\gamma$ estimate a probability of size $n^{-\alpha}$ up to an additive error $\gamma n^{-\alpha}$ then if is sufficient to estimate a probability of size $\lambda_1 n^{-\alpha}$ up to an additive error of $\gamma\sqrt{\lambda_1}n^{-\alpha}$, ie, $\gamma_2 \leq \frac{\gamma}{\sqrt{\lambda_1}}$. Hence we get

$$
\begin{aligned}
&|(2\gamma_2 + \gamma_2^2)(\lambda_1\lambda_2)^2| + |2(\lambda_1\lambda_2)(\gamma_1 + \gamma_2 + \gamma_1\gamma_2)| + |(2\gamma_1 + \gamma_1^2)| + |\epsilon(1+\epsilon)| \\
&\leq (2\gamma'\lambda_1^{3/2}\lambda_2^2 + \gamma'^2\lambda_1\lambda_2^2 + 2\gamma'\lambda_1\lambda_2 + 2\gamma'\sqrt{\lambda_1}\lambda_2 + 2\gamma^2\sqrt{\lambda_1}\lambda_2 + 3\gamma' + 2\gamma'^2 \\
&\leq (2\gamma'(1+\gamma')^2 + \gamma'^2(1+\gamma')^2 + 2\gamma'(1+\gamma') + 2\gamma'(1+\gamma') + 2\gamma'^2(1+\gamma') + 3\gamma' + 2\gamma'^2 \\
&\leq \gamma/10
\end{aligned}
$$

for small $\gamma$. Hence we again get that $|\frac{(\tilde{p}_i-\tilde{q}_i)^2}{\tilde{p}_i+\tilde{q}_i} - \frac{(p_i-q_i)^2}{p_i+q_i}| \leq p_i\gamma/10$. □

*Proof of Lemma 6.* Let $A = \sum_{i \in S} \frac{(p_i-q_i)^2}{p_i+q_i}$ and $B = \sum_{i \notin S} \frac{(p_i-q_i)^2}{p_i+q_i}$. By Lemma 5 we estimate $A$ with an additive error of $\epsilon/10$. Now, $\max(||p'||_\infty, ||q'||_\infty) < n^{-\alpha}(1+\epsilon)$. If $\Delta(p,q) > \epsilon$ then either $A$ is bigger than $\epsilon/2$ or $B$ is bigger than $\epsilon/2$. If $A$ is bigger than $\epsilon/2$ then our estimate of $A$ is bigger than $\epsilon(1/2-2/10)$ and in which case $\sum_{i \in S} \frac{(\tilde{p}_i-\tilde{q}_i)^2}{\tilde{p}_i+\tilde{q}_i} > \epsilon/10$ and we fail. Otherwise if $B$ is bigger than $\epsilon/2$ then $\ell_2(p',q') > \frac{\epsilon}{\sqrt{n}2}$ and the $\ell_2$ test fails since $\sqrt{n}\ell_2(p',q') \geq \epsilon/2$. Alternatively if $\Delta(p,q) < \epsilon^2/n^{1-\alpha}$ then $A < \epsilon^2/n^{1-\alpha}$ and we don't fail at the first hurdle. Also $B < \epsilon^2/n^{1-\alpha}$ implies the second test passes since $n^\alpha \ell_2^2(p',q') \leq \epsilon^2/n^{1-\alpha}$ and thus $\ell_2(p',q') \leq \frac{\epsilon}{2\sqrt{n}}$ (appealing to Eq. 1.) □

*Proof of Lemma 19.* We use ideas from the simulation in Theorem 25. We want to find $t = O(\epsilon^{-2}\log n)$ elements from the projection at random with replacement ($H^{-1}$ is constant). Let $MS$ be the multi-set of items which allowed us to conclude that the entropy is large. We keep the subsequent counts of all the elements in $MS$. We also look at the first $t$ elements of the projection (irrespective of membership in $MS$), denoted by Prefix and keep track of the new elements seen. At the end of the stream, we know the length $m$, and can now simulate the combined oracle algorithm using $MS$ and Prefix. The first element of Prefix is almost a random sample, with a corrective term that depends on the size of $MS$ and $m$. If there is a correction, then we choose a random element of $MS$. If we do not correct, we output the first element of Prefix, add it to $MS$ and shift the Prefix to the next element and repeat the process. □

*Proof of Theorem 21.* If there is no such $i'$, we are guaranteed (whp, with high probability) that the entropy $H$ of the projection satisfies $H \geq 1$ and the simulation is successful whp as well. If there is such an $i'$ then whp we decrease the mass of the projection by factor 2. This can repeat for at most $8\log m = O(\log n)$ steps ($m$ is the stream length) since after that the probability mass of the residual distribution is smaller than a factor $1/m^2$ of the original distribution. The contribution of these elements towards the entropy $H(p)$ of the original distribution is at most $\frac{2\log m + \log n}{m^2}$. But if there are more than one element in the distribution then the minimum entropy is $\log mm$, since worst case is when $m-1$ elements are the same and 1 element is different (follows from concavity). So we can ignore the contribution of these elements (note that after the first projection we are guaranteed that there are two distinct elements in the stream). □

*Proof of Lemma 14.* The RHS follows from the fact that for sufficiently large $m$, we have $(1-x/m) \geq e^{-(1+\frac{1}{\sqrt{m}})\frac{x}{m}}$. Therefore $Pr[\chi_{ij}] \leq 1 - \left(1 - \frac{(1+\epsilon)\cdot 2^j}{mt}\right)^{mp_i} \leq 1 - e^{-(1+\epsilon)^2 \frac{p_i 2^j}{t}} \leq (1+\epsilon)^2 \frac{p_i 2^j}{t}$ since $(1+\epsilon)^2 \frac{p_i 2^j}{t} \leq 1$. The LHS follows analogously, $Pr[\chi_{ij}] \geq 1 - \left(1 - \frac{2^j}{mt}\right)^{mp_i} \geq 1 - e^{-\frac{2^j p_i}{t}} \geq (1-\frac{1}{e})\frac{2^j p_i}{t}$. □

*Proof of Lemma 15.* Consider $\mathbf{E}\left[\sum_j \frac{f_j}{2^j}\right]$

$$\begin{aligned}
\mathbf{E}\left[\sum_j \frac{f_j}{2^j}\right] &= \sum_j \mathbf{E}\left[\frac{f_j}{2^j}\right] = \sum_j \frac{1}{2^j} \sum_i Pr[\chi_{ij}] \leq \sum_i \sum_{j=1}^\infty \left[1 - \left(1 - \frac{(1+\epsilon) \cdot 2^j}{mt}\right)^{mp_i}\right] \frac{1}{2^j} \\
&\leq \sum_i \left\{ \sum_{j=1}^{\lfloor \log \frac{1}{p_i} \rfloor} \left[1 - \left(1 - \frac{(1+\epsilon) \cdot 2^j}{mt}\right)^{mp_i}\right] \frac{1}{2^j} + \sum_{j=\lfloor \log \frac{1}{p_i} \rfloor + 1}^\infty \frac{1}{2^j} \right\} \\
&\leq \sum_i \left\{ \sum_{j=1}^{\lfloor \log \frac{1}{p_i} \rfloor} \left[1 - \left(1 - \frac{(1+\epsilon) \cdot 2^j}{mt}\right)^{mp_i}\right] \frac{1}{2^j} + 2p_i \right\} \\
&\leq \sum_i 2p_i + \sum_{j=1}^{\lfloor \log \frac{1}{p_i} \rfloor} (1+\epsilon)^2 \frac{p_i 2^j}{t} \frac{1}{2^j} \leq \sum_i 2p_i + (1+\epsilon)^2 \frac{p_i \log \frac{1}{p_i}}{t} \leq \frac{1}{t}(1+\epsilon)^2 H + 2
\end{aligned}$$

Similarly,

$$\begin{aligned}
\mathbf{E}\left[\sum_j \frac{f_j}{2^j}\right] &= \sum_{i,j} \frac{1}{2^j} Pr[\chi_{ij}] \geq \sum_i \sum_{j=1}^\infty \left[1 - \left(1 - \frac{2^j}{mt}\right)^{mp_i}\right] \frac{1}{2^j} \geq \sum_i \sum_{j=1}^{\lfloor \log \frac{1}{p_i} \rfloor} \left[1 - \left(1 - \frac{2^j}{mt}\right)^{mp_i}\right] \frac{1}{2^j} \\
&\geq \sum_i \sum_{j=1}^{\lfloor \log \frac{1}{p_i} \rfloor} (1 - \frac{1}{e}) \frac{2^j p_i}{t 2^j} \geq \frac{1}{t}(1 - \frac{1}{e}) \sum_i p_i (\log \frac{1}{p_i} - 1) = \frac{1}{t}(1 - \frac{1}{e})(H - \sum_i p_i)
\end{aligned}$$

which proves the lemma. □

*Proof of Lemma 24.* Note that `sample` does not take a parameter and therefore only the number of samples we make can depend on the outcome of probes we may do. However, we know that there can be at most $t$ samples taken. Hence if we request $t$ samples initially we can assume that we do not need to do any further sampling. (Note that we have at most doubled the oracle calls). Let $S$ be the set of $i$'s seen as samples. Obviously taking more samples than were taken by $\mathcal{A}$ can not be detrimental when it comes to correctness. Then, for each value $i \in S$ we perform $\text{probe}(p, i)$. This only adds $t$ queries to the complexity.

We now have a randomized algorithm that takes as input the outcome from our $t$ samples and the value $p_i$ for all $i \in S$ and performs a series of probes $\text{probe}(p, j)$ (wlog $j \notin S$). Note that since all samples have already been made, this phase of the computation can be viewed as a randomized decision tree. At each node, the algorithm either tosses a random coin or makes a query $\text{probe}(p, j)$, and moves to one of the children of this node based on all values seen thus far (we assume the algorithm makes no decision based on the value of $i$, the so-called *variable-oblivious* model).

We now invoke a simplification of a lemma due to Bar-Yossef et al [10, 6], as follows:

**Lemma( [6] Lemma 4.17)** *Let T be a randomized decision tree that computes an $\epsilon$-approximation to f (where f is symmetric) with $u$ queries in the worst case and $u_E$ queries in the expected case (with the expectation taken over the random choices used by T). Then there is an oblivious decision tree $(T, W_u)$ (where $W_u$ is the uniform-without-replacement distribution) of worst-case query complexity $u$ and expected query complexity $u_E$ that computes an $\epsilon$-approximation to $f$.*

Our strategy is therefore as follows: Rewrite the randomized decision tree that generates $t$ probes to be an oblivious decision tree. In such a tree, all queries can be decided in advance and we now have an algorithm of the desired canonical form. □

*Proof of Theorem 25.* Consider the following streaming algorithm that uses $O(t \log n \log m)$ spaces. We store the first $t$ items in the data stream, $\texttt{Prefix}_t = (\langle p, i_1, + \rangle, \langle p, i_2, + \rangle, \ldots \langle p, i_t, + \rangle)$. Now for each $i \in S_w = \{i_1, i_2, \ldots i_t\}$ we set up a counter that will be used to maintain as exact count of the frequency of $i$. We now chose $t$ values, $S'_w$ from $k \in [n] \setminus S_w$ uniformly at random (without replacement) and set up a counter for each of these $t$ values. We also maintain a counter to estimate the length of the stream $m$. At the end of the data stream we claim that we can simulate the oracles calls made by any canonical algorithm. The only difficulty in establishing this claim is showing that we can use the $\texttt{Prefix}_t$ to simulate the generative oracle. Ideally we would like to claim that we can just return $i_j$ on the $j$th query to the generative oracle. This is however not the case. Eg. if the frequency of the element 10 is $f_{10}$ and the length of the stream is $m$ then, if $i_1 = 10$ then the probability that $i_2 = 10$ is $\frac{f_{10}-1}{m} \neq \frac{f_{10}}{m} = p_{10}$ whereas, in the generative oracle, the event that the second sample is 10 is independent of the first sample. We get around this dependence in the random stream model by doing the following: on the $j$th generative oracle call we output $i_j$ with probability $(m - j + 1)/m$ and otherwise output $i_{j'}$ where $j'$ is chosen uniformly at random from $[j-1]$. We thus can emulate the generative oracle calls made by $\mathcal{A}$. The emulation of the evaluative oracle calls, ie. the probes performed by $\mathcal{A}$, can also be emulated because for each $i_j$ we have maintained counters that give us $p_{i_j}$ and for each $k \in S'_w$ we know $p_k$.

For the two pass regular stream algorithm things are simpler. In the first pass we generate our random sample (with replacement) using standard techniques. In the second pass we count the exact frequencies of the relevant $i$. □

## B  The Large–Small algorithm for Entropy

The algorithm we will use can be described as a block-by-block process or a much more elegant simplification reminiscent of online facility location algorithm of Meyerson [32] in the context of stream clustering. There are also some similarities with the count/count-min sketches of [14, 16]. Our algorithm will "track" a few items, i.e., maintain explicit counters for them. Let $1 > \alpha > 0$ be a constant to be decided later. Let $H$ be the true entropy and $p_i$ the true probability of $i$. Assume $m > n \geq 3$. The algorithm is presented in Figure 5.

**Lemma 26.** *Let the true entropy be $H$, then $\hat{H} \geq H$.*

*Proof.* The fact that we never underestimate the entropy follows from the fact that we never *overestimate* the probabilities. The fact that we distribute the "residual" properties among the untracked elements in uniform way, ensures that the lemma holds. We can also derive an algebraic proof, but the information theoretic proof is simpler. □

In the rest of the subsection we will show that we do not *overestimate* the entropy, i.e., $H \not\ll \hat{H}$. The proof that we correctly track the "large" probability elements is straightforward.

**Claim 27.** *If $2\tilde{m} \geq m \geq \tilde{m}$ (for appropriate guess $\tilde{m}$) and $p_i \geq n^{-\alpha}/2$ then $i$ is tracked with probability $1 - n^{-2}$, furthermore, $|c(i)/m - p_i| \leq \epsilon n^{-\alpha}$.*

Notice that the size of $\bar{S}$ is at most $n^\alpha$, hence we have,

**Corollary 28.** *Let $S = \{i|p_i \leq n^{-\alpha}\}$ and $w = \sum_{i \in S} p_i$. Then whp., $\hat{S} = S$ and $\hat{w} \geq w \geq \hat{w} - \epsilon$.*

The proof of the next claim is not too difficult, but there is a small issue that $x \log \frac{1}{x}$ is a bitonic (increasing and then decreasing) sequence with maxima at $1/e$.

> **Algorithm** *Entropy-Estimation*
> 1. Guess $\tilde{m} = 1, 2, 4, \ldots 2^i$ where $\tilde{m}$ is roughly the number of insertions.
> 2. For each guess of $\tilde{m}$:
>    - Initially $m = 0$.
>    - On new input $\langle i, + \rangle$, increase $m$. If $m > 2\tilde{m}$ abort process.
>    - If $i$ is being "tracked", increase counter of $i$. Else with prob. $\frac{\log n}{\epsilon n^\alpha m}$ decide to start tracking $i$.
>    - Keep checking if at least 2 elements are seen, i.e., the entropy is non-zero.
> 3. At the end of input focus on the guess where $2\tilde{m} \geq m \geq \tilde{m}$, since we know $m$.
> 4. Define the collection of tracked objects $\langle i, c(i) \rangle$ as the "signature" of the stream.
> 5. If $c(i)$ is the count of $i$, let $\bar{S}$ be the set of tracked elements with $c(i) \geq n^\alpha$. (The choice of the notation implies $S$ is the set of "small" probability elements.) Let $\hat{S}$ be the complement of $\bar{S}$.
> 6. For $i \in \bar{S}$ define $h_i = \frac{c(i)}{m} \log \frac{m}{c(i)}$
> 7. Let $\hat{w} = 1 - \sum_{i \in \bar{S}} c(i)/m$. Define $\hat{H}_{\hat{S}} = \hat{w} \log \frac{n}{\hat{w}}$. We will assume $n > 3$.
> 8. Output $\hat{H} = \hat{H}_{\hat{S}} + \sum_{i \in \bar{S}} h_i$.
>
> Figure 5: The Large-Small Algorithm for Estimating Entropy in a Data Stream

**Lemma 29.** *Whp.* $\sum_{i \in \bar{S}} h(i) \leq (1+\epsilon) \sum_{i \in \bar{S}} p_i \log \frac{1}{p_i} + 2\epsilon n^{-\alpha}$.

*Proof.* For all probabilities *less than* $1/e$ we can at most suffer a loss of $(1+\epsilon)$ factor since we underestimate the probability. But for $p_i \geq 1/e$ the derivative is bounded and at most $-1$. This applies to at most two elements. Observe that the worst case arises when $\sum_{i \in \bar{S}} p_i \log \frac{1}{p_i}$ (which is a convex function) is as low as possible. This means that the worst case is when exactly one $p_i \sim 1$. But the derivative of $p_i \log \frac{1}{p_i}$ is bounded and $|p_i \log \frac{1}{p_i} - h(i)|$ is at most $2(p_i - c(i)/m)$ which is bounded by Claim 27. □

Now let us focus on the small probability elements.

**Lemma 30.** *Let* $H_S = \sum_{i \in S} p_i \log \frac{1}{p_i}$ *(the true contribution of small elements) then* $\hat{H}_{\hat{S}} - \epsilon \log \frac{n}{\epsilon} \leq \frac{1}{\alpha}(1 + \frac{\log(1/\epsilon)}{\log n}) H_S$

*Proof.* For each $i \in S$, $p_i \leq n^{-\alpha}$. $H_S$ is minimized if given $p_i, p_j$ we set $p'_i = p_i + p_j$ and $p'_j = 0$ (by convexity). Now to respect $p_i \leq n^{-\alpha}$, the vector $\{p_i\}$ must have $p_i = n^{-\alpha}$ or $p_i = 0$, for all $p_i$ except 1. This last $i$ takes care of the excess, i.e., $p'_i = w - \lfloor \frac{w}{n^{-\alpha}} \rfloor n^{-\alpha}$. Therefore

$$H_S \geq \left\lfloor \frac{w}{n^{-\alpha}} \right\rfloor \log \frac{1}{n^{-\alpha}} + (w - \lfloor \frac{w}{n^{-\alpha}} \rfloor n^{-\alpha}) \log \frac{1}{w - \lfloor \frac{w}{n^{-\alpha}} \rfloor n^{-\alpha}}$$

We immediately have two subcases:

**Case $w < n^{-\alpha}$:** In this case $H_S \geq w \log \frac{1}{w}$ and

$$\hat{H}_{\hat{S}} - \epsilon \log \frac{n}{\epsilon} \leq (w + \epsilon) \log \frac{n}{w + \epsilon} - \epsilon \log \frac{n}{\epsilon} \leq w \log \frac{n}{\epsilon} \leq H_S \frac{\log n + \log \frac{1}{\epsilon}}{-\log w} \leq H_S \frac{\log n + \log \frac{1}{\epsilon}}{\alpha \log n}$$

**Case $w \geq n^{-\alpha}$:** In this case $H_S \geq \frac{w}{2n^{-\alpha}} \log n^\alpha$ and therefore,

$$\hat{H}_{\hat{S}} - \epsilon \log \frac{n}{\epsilon} \leq w \log \frac{n}{\epsilon} \leq 2 H_S n^{-\alpha} \frac{\log n + \log \frac{1}{\epsilon}}{\alpha \log n} \leq H_S \frac{\log n + \log \frac{1}{\epsilon}}{\alpha \log n} \quad \text{(for large enough } n\text{)}$$

Thus the first case dominates and the lemma is true. □

Putting together the Lemmas 29 & 30 we get:

**Theorem 31.** *We can estimate the entropy upto a factor $\frac{1}{\alpha}(1 + \frac{\log \frac{1}{\epsilon}}{\log n})$ and an additive term $\epsilon(\log \frac{n}{\epsilon} + n^{-\alpha})$ in space $O(\epsilon^{-1} n^\alpha \log n)$.*

We will assume that $H$ is bounded away from 0. Let $\epsilon' = \frac{\alpha(\epsilon \log(n/\epsilon) + n^{-\alpha})}{H} + \frac{\log 1/\epsilon}{\log n}$ and let $f(\epsilon', H)$ be equal to the $\epsilon$ satisfying this expression.

**Theorem 18.** We can estimate the entropy $H$ upto a factor $\frac{1}{\alpha}(1 + \epsilon')$ using space $O(\frac{n^\alpha \log n}{f(\epsilon', H)})$.